\documentclass[aps,prd,superscriptaddress,preprint,showpacs,tightenlines,nofootinbib]{revtex4}
\usepackage{graphicx}
\usepackage{dcolumn}
\usepackage{bm}
\usepackage{color}
\usepackage{relsize}
\usepackage{graphicx} 
\usepackage{dcolumn}  
\usepackage{bm}       

\newcommand{\rmn}[1]{\uppercase \expandafter {\romannumeral #1}}

\def\de {\Delta E}
\def \kkp {D^+\to K^+ K^-\pi^+}

\def \msm {m_-^2}
\def \msp {m_+^2}
\def \mbc {m_{\rm BC}}

\def\Dp {D^+}
\def \Dm {D^-}
\def\simlt{\mathrel{\lower2.5pt\vbox{\lineskip=0pt\baselineskip=0pt
          \hbox{$<$}\hbox{$\sim$}}}}
\def\babar{\mbox{\slshape B\kern-0.1em{\smaller A}\kern-0.1em
    B\kern-0.1em{\smaller A\kern-0.2em R }}}
\graphicspath{{eps/}}
\begin{document}

\preprint{CLNS 08/2036}       
\preprint{CLEO 08-19}         

\title{Search for $\boldmath CP$ Violation in the Dalitz-Plot Analysis of $\boldmath D^\pm\to K^+K^-\pi^\pm$}
\author{P.~Rubin}
\affiliation{George Mason University, Fairfax, Virginia 22030, USA}
\author{B.~I.~Eisenstein}
\author{I.~Karliner}
\author{S.~Mehrabyan}
\author{N.~Lowrey}
\author{M.~Selen}
\author{E.~J.~White}
\author{J.~Wiss}
\affiliation{University of Illinois, Urbana-Champaign, Illinois
61801, USA}
\author{R.~E.~Mitchell}
\author{M.~R.~Shepherd}
\affiliation{Indiana University, Bloomington, Indiana 47405, USA }
% ...
\collaboration{CLEO Collaboration} \noaffiliation

\begin{abstract}

We report on a search for $CP$ asymmetry in the singly
Cabibbo-suppressed decay $\kkp$ using a data sample of 818 pb$^{-1}$
accumulated with the CLEO-c detector on the $\psi(3770)$ resonance.
A Dalitz-plot analysis is used to determine the amplitudes of the
intermediate states. We find no evidence for $CP$ violation either
in specific two-body amplitudes or integrated over the entire phase
space. The $CP$
asymmetry 
in the latter case is measured to be
$(-0.03\pm0.84\pm 0.29)\%$.
\end{abstract}


\maketitle

$D$-meson decays are predicted in the Standard Model (SM) to exhibit
$CP$-violating charge asymmetries smaller than ${\mathcal
O}(10^{-3})$ \cite{SM}. Measurement of a $CP$ asymmetry in the $D$
system with higher rate would clearly signal new physics (NP)
\cite{bigi,nir}. Singly Cabibbo-suppressed (SCS) decays via $c\to
u\bar{q}q$ transitions are sensitive to NP contributions to the
$\Delta C=1$ penguin process. Interestingly, such processes do not
contribute to either the Cabibbo-favored ($c\to s\bar{d}u$) or the
doubly Cabibbo-suppressed ($c\to d\bar{s}u$) decays. Direct $CP$
violation in SCS decays could arise from interference between tree
and penguin  processes. A non-zero $CP$ asymmetry can occur if
there is both a strong and weak phase difference between
the tree and penguin processes. 
In charged $D$-meson decays, mixing effects are absent, allowing us
to probe direct $CP$ violation and consequently NP.

Weak decays of $D$ mesons are expected to be dominated by quasi
two-body decays with resonant intermediate states. Dalitz-plot
analysis techniques
can be used to explore the resonant substructure. 
The intermediate structures of $\kkp$ decay were studied by E687
\cite{e687} with a Dalitz-plot analysis and by FOCUS \cite{focus}
with a non-parametric technique. \babar searched for direct $CP$
asymmetries in this mode using a counting method \cite{babar}. Using
281 pb$^{-1}$ of data, CLEO previously measured the absolute
hadronic branching fractions and the $CP$ asymmetries of
Cabibbo-favored $D$-meson decay modes and the phase-space integrated
asymmetry in the $K^+K^-\pi^+$ mode we study here \cite{cleo281}.
The previous investigations of this decay were either limited by
statistics, and did not search for $CP$ violation, or did not study
the resonant substructure.

We present the results of a search for direct $CP$ asymmetry in the
decay $D^\pm \to K^+K^-\pi^\pm$. This includes a study of the
integrated decay rate, as well as decays through various
intermediate states. We perform the present analysis on 818
pb$^{-1}$ of $e^+e^-$ collision data collected at a center-of-mass
energy of 3774 MeV with the CLEO-c detector \cite{det1,det2,det3} at
the Cornell Electron Storage Ring (CESR). The CLEO-c detector is a
general purpose solenoidal detector that includes a tracking system
for measuring momentum and specific ionization ($dE/dx$) of charged
particles, a Ring Imaging Cherenkov detector (RICH) to aid in
particle identification, and a CsI calorimeter for detection of
electromagnetic showers. 

\begin{figure}[!hbtp]
\includegraphics[width=0.8\textwidth]{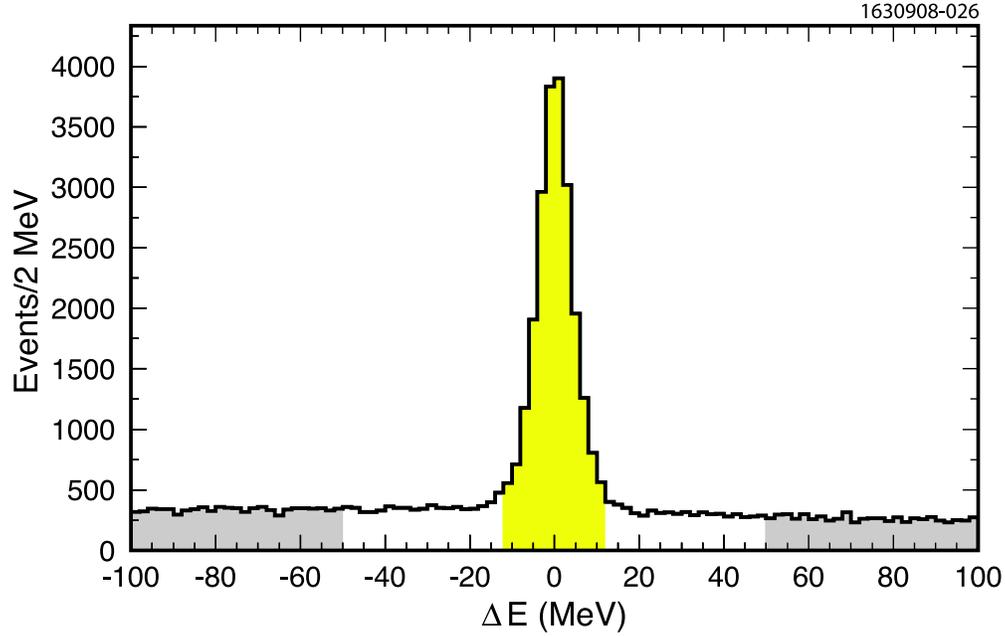}
\caption{The $\de$ distributions. Signal ($|\Delta E|<12$~MeV) and
sidebands (50~MeV$<|\de|<$ 100~MeV) regions are shown. \label{de}}
\end{figure}

We reconstruct $D^+\to K^+ K^-\pi^+$, and the charge-conjugate mode
$D^-\to K^+ K^-\pi^-$. (Charge-conjugate modes are included
throughout this report unless noted otherwise.) The event
reconstruction criteria are the same as that used in
Ref.~\cite{cleo281}.  Charged tracks are required to be well
measured and to satisfy criteria based on the track fit quality.
They must also be consistent with coming from the interaction
point in three dimensions. Pions and kaons are identified 
using $dE/dx$ and RICH information, when available. If either
$dE/dx$ or RICH information (or both) is missing we still use the
track in the analysis. Detail can be found in Ref.~\cite{cleo281}.
We define two signal variables:
\begin{equation}
\Delta E \equiv \sum_iE_i-E_{\rm beam}
\end{equation}
and 
\begin{equation}
m_{\rm BC} \equiv \sqrt{E_{\rm beam}^2-|\sum_i \mathbf{p}_i|^2}\,,
\end{equation}
where $E_i$ and $\mathbf{p}_i$ are the energy and momentum of each
$D$ decay product, and $E_{\rm beam}$ is the energy of one of the
beams. For a correct combination of particles, $\de$ should be
consistent with zero, and $\mbc$ should be consistent with the $D^+$
mass. Fig. \ref{de} shows $\de$ distribution of data. We select
candidates that have $\de$ within $\pm12$ MeV of zero, corresponding
to 2.5 standard deviations ($\sigma$). If in any event there are
multiple candidates satisfying the $\de$ criterion using entirely
separate combinations of tracks, we accept
all of these candidates. 
Otherwise if there are multiple candidates sharing tracks we keep
only the combination with the smallest $|\Delta E|$.

\begin{figure}[!hbtp]
\includegraphics[width=0.8\textwidth]{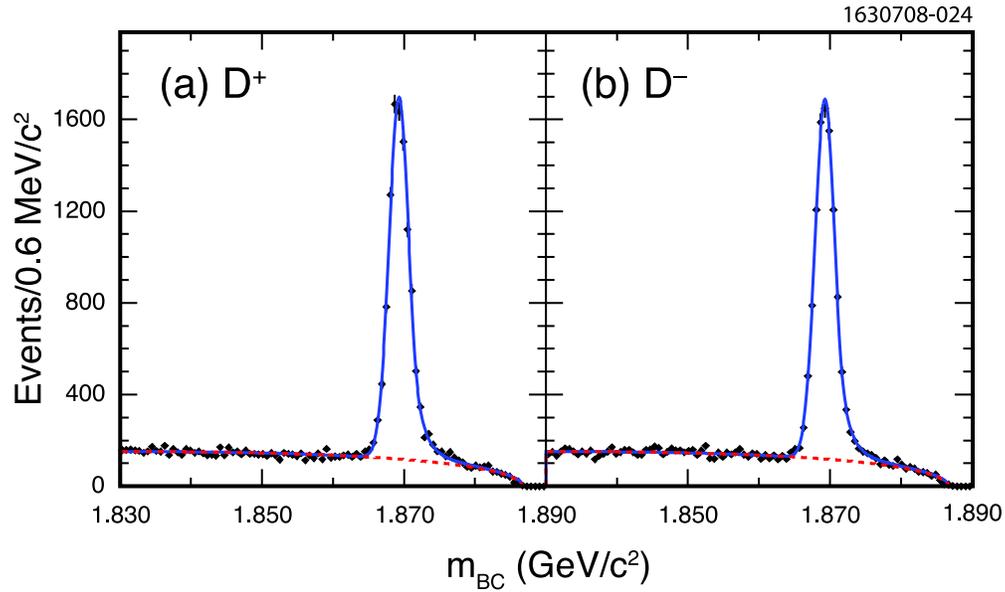}
\caption{The $\mbc$ distributions for (a) $D^+$ and (b) $D^-$
candidates. The solid curves show the fits to the data (points with
error bars), while the dashed curves indicate the background.
\label{mbc}}
\end{figure}

To determine the signal yields of the $D^+$ and $D^-$ samples, we
simultaneously fit the $\mbc$ distributions from the samples and
require they have the same signal shape. For the signal, we use a
Crystal Ball line shape function \cite{cb}, whose parameters are
allowed to float. For the background, an ARGUS function \cite{argus}
is used with shape parameters determined from the events in the
$\de$ sideband (50 MeV $<|\de|<$ 100 MeV). We find $9757\pm116$
$D^+$ and $9701\pm115$ $D^-$. Figure \ref{mbc} shows the $m_{\rm
BC}$ distributions of $D^+$ and $D^-$ samples with fit functions
superimposed; the total $\chi^2$ is 241 for 180 degrees of freedom
(d.o.f.).

We obtain the efficiency from a GEANT-based signal Monte Carlo (MC)
simulation of the detector. The signal MC requires one of the two
$D$ mesons in an event to decay in accordance with all known modes
and the other one to decay to the signal mode.
For the signal $D$ meson, we generate events that uniformly populate
phase space.  The average efficiency, accounting  for a non-uniform
population density of data, is calculated as follows. The Dalitz
plot of the data is first divided into 16 bins that are
approximately equally populated. The signal yields are obtained from
the $m_{\rm BC}$ fits bin by bin and the corresponding efficiencies
are calculated from the MC. The average efficiency is the sum of the
yields divided by the sum of the efficiency-corrected yields. We
find the efficiencies $\epsilon^{\pm}$ for the $D^\pm$ decays are
$(44.13\pm0.15)\%$ and $(43.85\pm0.15)\%$, respectively. The $CP$
asymmetry, defined as
\begin{equation}
A_{CP}=\frac{N^+/\epsilon^+ - N^-/\epsilon^-}{N^+/\epsilon^+ +
N^-/\epsilon^-},\end{equation} where $N^\pm$ are the measured
$D^\pm$ yields, is measured as \begin{equation}
A_{CP}=(-0.03\pm0.84\pm0.29)\%~. \label{acp_yield}\end{equation}

\begin{table*}[btp]
\caption{\label{dpfit}Fit results for three models with different
$S$-wave parameterizations. The $K^-\pi^+$ $S$-wave contains
contributions from $\overline{K}^*_0(1430)^0$ and a nonresonant term
in fit A, from $\overline{K}^*_0(1430)^0$ and $\kappa(800)$ in fit
B, and from the LASS amplitude in fit C. The errors are statistical,
experimental systematic, and decay-model systematic, respectively. }
\begin{tabular}{lccc}\hline\hline
 &Magnitude &Phase ($^\circ$)&Fit Fraction (\%)\\\hline
 \multicolumn{4}{c}
 {Fit A [$\chi^2/$d.o.f. = 898/708]}\\
$\overline{K}^{*0}$& 1(fixed)&0(fixed)&$25.0\pm0.6_{-0.3-1.2}^{+0.4+0.2} $\\
$\overline{K}_0^*(1430)^0$&$3.7\pm0.5_{-0.1-1.0}^{+0.5+1.0}$&$73\pm9_{-6-38}^{+6+15}$&
$12.4\pm3.3_{-0.7-5.8}^{+3.4+7.3}$ \\
$\phi$&$1.189\pm0.015_{-0.011-0.010}^{+0.000+0.028}$&$-179\pm4_{-1-5}^{+3+13}$&
$28.1\pm0.6_{-0.3-0.4}^{+0.1+0.2}$\\
$a_0(1450)^0$&$1.72\pm0.10_{-0.11-0.28}^{+0.11+0.81}$&$123\pm3_{-1-15}^{+1+9}$&
$5.9\pm0.7_{-0.6-1.8}^{+0.7+6.7}$\\
$\phi(1680)$&$1.9\pm0.2_{-0.1-0.7}^{+0.0+1.3}$&$-52\pm8_{-5-26}^{+0+10}$&
$0.51\pm0.11_{-0.04-0.12}^{+0.02+0.85}$\\
$\overline{K}_2^*(1430)^0$&$6.4\pm0.9_{-0.4-3.6}^{+0.5+1.9}$&$150\pm6_{-0-13}^{+1+28}$&
$1.2\pm0.3_{-0.1-0.6}^{+0.2+0.8}$\\
NR&$5.1\pm0.3_{-0.3-0.2}^{+0.0+0.6}$&$53\pm7_{-5-11}^{+1+18}$&$14.7\pm1.8_{-1.6-1.5}^{+0.2+3.9}$\\
\multicolumn{4}{c} {Total Fit Fraction  = $(88.7\pm2.9)\%$}\\
\hline
 \multicolumn{4}{c}{Fit B [$\chi^2/$d.o.f. = 895/708]}\\
$\overline{K}^{*0}$& 1(fixed)&0(fixed)&$25.7\pm0.5_{-0.3-1.2}^{+0.4+0.1}$\\
$\overline{K}_0^*(1430)^0$&$4.56\pm0.13_{-0.01-0.39}^{+0.10+0.42}$&$70\pm6_{-6-23}^{+1+16}$&
$18.8\pm1.2_{-0.1-3.4}^{+0.6+3.2}$ \\
$\phi$&$1.166\pm0.015_{-0.009-0.009}^{+0.001+0.025}$&$-163\pm3_{-1-5}^{+1+14}$&
$27.8\pm0.4_{-0.3-0.4}^{+0.1+0.2}$\\
$a_0(1450)^0$&$1.50\pm0.10_{-0.06-0.33}^{+0.09+0.92}$&$116\pm2_{-1-14}^{+1+7}$&
$4.6\pm0.6_{-0.3-1.8}^{+0.5+7.2}$\\
$\phi(1680)$&$1.86\pm0.20_{-0.08-0.77}^{+0.02+0.62}$&$-112\pm6_{-4-12}^{+3+19}$&
$0.51\pm0.11_{-0.04-0.15}^{+0.01+0.37}$\\
$\overline{K}_2^*(1430)^0$&$7.6\pm0.8_{-0.6-4.8}^{+0.5+2.4}$&$171\pm4_{-2-11}^{+0+24}$&
$1.7\pm0.4_{-0.2-0.7}^{+0.3+1.2}$\\
$\kappa$(800)&$2.30\pm0.13_{-0.11-0.29}^{+0.01+0.52}$&$-87\pm6_{-3-10}^{+2+15}$&$7.0\pm0.8_{-0.6-1.9}^{+0.0+3.5}$\\
\multicolumn{4}{c} {Total Fit Fraction  = $(86.1\pm1.1)\%$}\\\hline
 \multicolumn{4}{c}{Fit C  [$\chi^2/$d.o.f. = 912/710]}\\
$\overline{K}^{*0}$& 1(fixed)&0(fixed)&$25.3\pm0.5_{-0.4-0.7}^{+0.2+0.2}$\\
LASS&$3.81\pm0.06_{-0.05-0.46}^{+0.05+0.13}$&$25.1\pm2_{-2-5}^{+1+6}$&$40.6\pm0.8_{-0.5-9.1}^{+0.4+1.6}$\\
$\phi$&$1.193\pm0.015_{-0.010-0.011}^{+0.003+0.021}$&$-176\pm2_{-2-8}^{+0+8}$&
$28.6\pm0.4_{-0.3-0.5}^{+0.2+0.2}$\\
$a_0(1450)^0$&$1.73\pm0.07_{-0.03-0.38}^{+0.14+0.68}$&$122\pm2_{-1-10}^{+1+8}$&
$6.0\pm0.4_{-0.2-2.4}^{+0.9+5.5}$\\
$\phi(1680)$&$1.71\pm0.16_{-0.02-0.77}^{+0.02+0.41}$&$-72\pm8_{-2-22}^{+2+10}$&
$0.42\pm0.08_{-0.01-0.16}^{+0.02+0.19}$\\
$\overline{K}_2^*(1430)^0$&$4.9\pm0.7_{-0.4-2.3}^{+0.1+2.2}$&$146\pm9_{-7-11}^{+0+34}$&
$0.7\pm0.2_{-0.1-0.3}^{+0.0+0.7}$\\
\multicolumn{4}{c} {Total Fit Fraction  = $(101.5\pm0.8)\%$}\\
 \hline\hline
\end{tabular}
\end{table*}

For the Dalitz-plot analysis, we consider the events from the signal
box ($|\de|<$12 MeV and $|\mbc-m_{D^+}|<4.5$ MeV/$c^2$)
corresponding to a 2.5$\sigma$ range in each variable. The signal
purity is $(84.26\pm0.10)\%$ obtained from the $\mbc$ fit. The
$K^+K^-\pi^+$ Dalitz-plot distribution is parameterized using the
isobar model formalism
described in Ref.~\cite{kopp}. The decay amplitude as a function of
Dalitz-plot variables is expressed as a sum of two-body decay matrix
elements,
\begin{equation}
{\cal M}(m_+^2,m_-^2)=\sum_r a_re^{i\delta_r}{\cal
A}_r(m_+^2,m_-^2),
\end{equation}
where each term is parameterized with a magnitude $a_r$ and a phase
$\delta_r$ for the intermediate resonance $r$, and $r$ ranges over
all resonances. We choose $m_+^2=m^2_{K^+\pi^+}$ and
$m_-^2=m^2_{K^-\pi^+}$ as the two independent Dalitz-plot variables.
The partial amplitude ${\cal A}_r(m_+^2,m_-^2)$ is parameterized
using the Breit-Wigner shape with Blatt-Weisskopf form factors in
the $D$ meson and intermediate resonance vertices \cite{bwform}, and
angular dependence taken into account \cite{kopp}.

We use an unbinned maximum likelihood fit which maximizes the
function
\begin{equation}
{\cal F}= \sum_{i=1}^N 2 \ln {\cal
L}(m_{+,i}^2,m_{-,i}^2)-\left(\frac{f-f_0}{\sigma_{f}}\right)^2,
\end{equation}
where the index $i$ runs over all $N$ events. The last term is used
to constrain the signal fraction $f$ to be the value $f_0$ within
its error $\sigma_f$ obtained from the $m_{\rm BC}$ fit. The first
term contains the likelihood function
\begin{equation}
{\cal L}(\msp,\msm)=f \frac{\varepsilon(\msp,\msm)|{\cal M}|^2}{
{\cal N}_{\rm sig}}+(1-f)\frac{F_{\rm bg}(\msp,\msm)}{{\cal N}_{\rm
bg}},\label{lk}
\end{equation}
where \begin{equation} {\cal N}_{\rm sig}=\displaystyle \int
\varepsilon(\msp,\msm)|{\cal M}|^2\,d{\msp} d{\msm}
 \end{equation}
and
\begin{equation}
{\cal N}_{\rm bg} =\displaystyle \int F_{\rm bg}(\msp,\msm)\,d{\msp}
d{\msm}
\end{equation}
are the normalization factors, and $\varepsilon(\msp,\msm)$ and
$F_{\rm bg}(\msp,\msm)$ are efficiency and background functions. The
fit parameters are $a_r$, $\phi_r$ and $f$.

We determine the efficiency $\varepsilon(\msp,\msm)$ using the same
signal MC sample described before. The efficiency function is
parameterized by a cubic polynomial in ($m_+^2$, $m_-^2$) multiplied
by threshold factors $T(m^2_{+\,max}-m_+^2;p_{xy}) \times
T(m^2_{-\,max}-m_-^2;p_{xy}) \times T(z_{max}-z;p_z)$, where
\begin{equation} T(x;p)=\left\{\begin{array}{l@{\quad,\quad}l}
\sin(px) & 0<px<\pi/2\\1 & {\rm otherwise}\end{array}\right.,
\end{equation}
$z \equiv m^2_{K^+K^-}$, $m^2_{\pm\,max}$ or $z_{max}$ is the
maximum value of $m^2_{\pm}$ or $z$ in this decay, $p_{xy}$ and
$p_z$ are the fit parameters. The threshold factors are used to
account for tracking inefficiency at the Dalitz-plot corners, where
one of three particles might be produced with very low momentum and
escape detection.

Figure \ref{mbc} shows that the background is significant. 
To construct a
model of the background shape $F_{\rm bg}(\msp,\msm)$, we select
events from the sideband region ($24<|\Delta E|<42$ MeV and $|m_{\rm
BC}-m_{D^+}|<9$ MeV/$c^2$). There are 12324 events, 
about 3.5 times the amount of background we estimate in the signal
region, which is dominated by random combinations of unrelated
tracks. Although the background includes $\phi$ and $K^*$ mesons
combined with random tracks, these events will not interfere with
each other. Thus the shape is parameterized by a two-dimensional
quadratic polynomial with terms representing non-coherent
contributions from $\phi$ and $K^*$ meson decays, multiplied by the
threshold factors.

We consider fifteen intermediate states, $\phi\pi^+$,
$\phi(1680)\pi^+$, $\overline{K}^{*0}K^+$,
$\overline{K}_0^*(1430)^{0}K^+$, $\overline{K}^*(1410)^{0}K^+$,
$\overline{K}_2^*(1430)^{0}K^+$, $\kappa(800) K^+$, $f_0(980)
\pi^+$, $f_0(1370)\pi^+$, $f_0(1500)\pi^+$, $f_2(1270)\pi^+$,
$f_2^\prime(1525)\pi^+$, $a_0(980)^0\pi^+$, $a_0(1450)^0\pi^+$ and
$a_2(1320)^0\pi^+$, as well as a nonresonant (NR) contribution. The
parameters of the established resonances are taken from
Ref.~\cite{pdg}, except for the $f_0(980)$ which is taken from
Ref.~\cite{f0} and the $a_0(980)$ taken from Ref.~\cite{a0}. A
complex pole function is used to model the $\kappa(800)$ with pole
position at $s_\kappa=(0.71-i0.31)^2$ GeV$^2$ \cite{kappa}. The
nonresonant contribution is modeled as a uniform distribution over
the allowed phase space. For the $K^-\pi^+$ $S$-wave states in the
decays, we also consider the LASS amplitude as described in
Ref.~\cite{LASS,BABARLASS}, instead of a coherent sum of the states
$\overline{K}_0^*(1430)^{0}K^+$, $\kappa (800) K^+$ and the
nonresonant term.

This study is sensitive only to relative phases and magnitudes. The
mode $\overline{K}^{*0}K^+$ is assigned to have zero phase and unit
magnitude. We choose the same phase conventions for the intermediate
resonances as E687 \cite{e687} used.

We begin to fit the data by considering only the three components
$\overline{K}^{*0}$, $\phi$, and $\overline{K}_0^*(1430)^0$ and
obtain a result consistent with E687. To present a relative
goodness-of-fit estimator, we divide the Dalitz-plot region into
bins with dimensions 0.05 (GeV/$c^2)^2$ $\times$ 0.05 (GeV/$c^2)^2$
and calculate $\chi^2$ as
\begin{equation}
\chi^2=-2\sum_{i=1}^{721}n_i\ln \left(\frac{p_i}{n_i}\right),
\end{equation}
where $n_i$ ($p_i$) is the observed (expected) number of events in
the $i$th bin \cite{pdg}. We find $\chi^2=1292$ for $(721-5)$ d.o.f.
in the ``three resonances" fit, where 721 is the
number of valid bins inside the kinematically allowed region. 

Our twenty times larger statistics than E687 require a better model.
We determine which additional resonances to include by the following
procedure: starting from the three resonances and adding new
resonances one at a time, we choose the best additional one at each
iteration,  stopping when no additional resonances have fit
fractions (FF) more than $3 \sigma$ from zero. The fit
fraction is defined as 
\begin{equation}
{\rm FF_r} = \frac{\displaystyle \int  |a_r {\cal A}_r|^2\, dm_+^2
dm_-^2} {\displaystyle \int |{\cal M}
|^2\,dm_+^2 dm_-^2}.\label{FF}
\end{equation}

\begin{figure}[!hbtp]
\center
\includegraphics[width=0.9\textwidth]{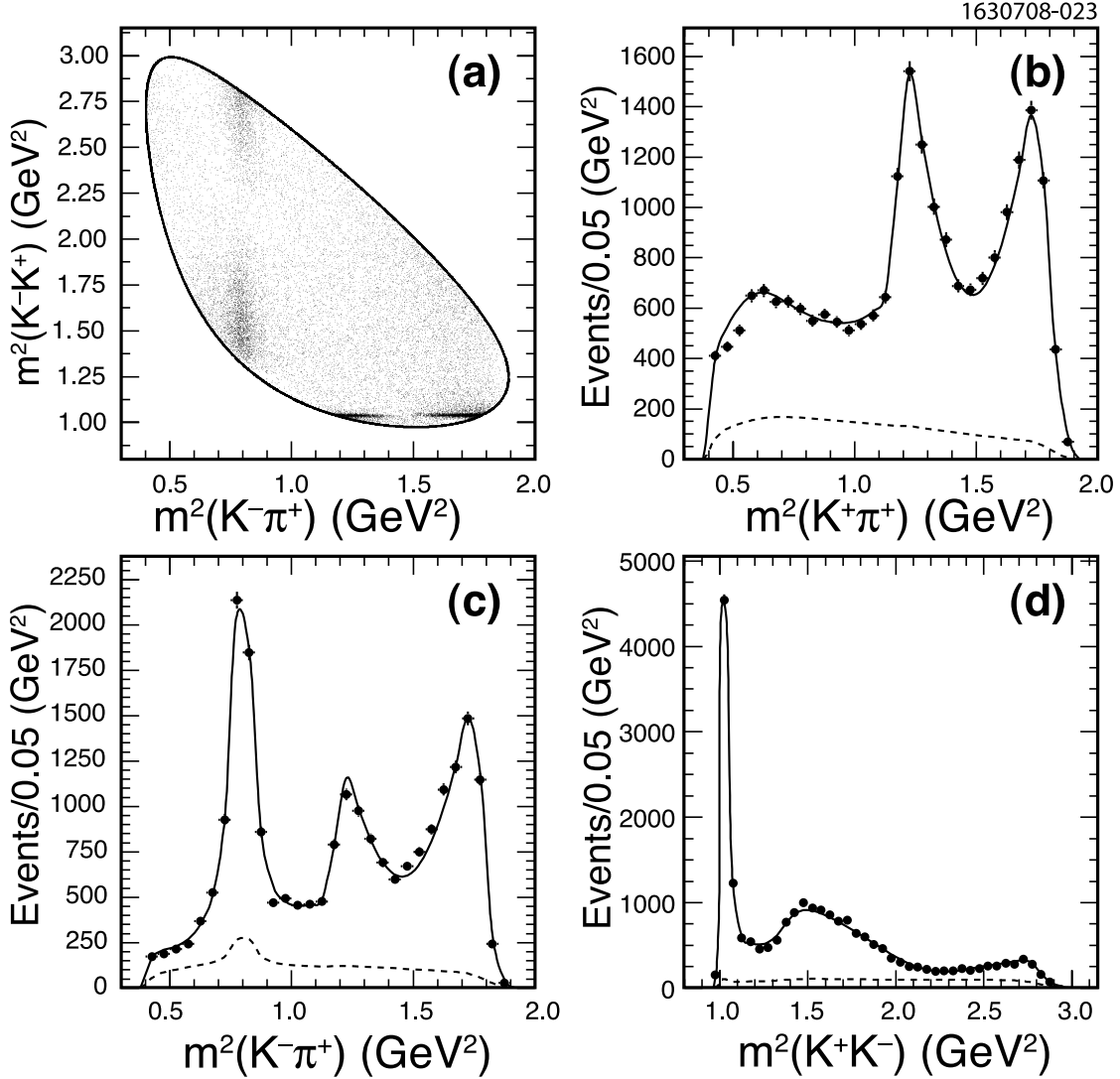}
\caption{\label{fig:lass}(a) The Dalitz plot for $\kkp$ candidates.
(b)-(d) Projections of the results of the fit B (line) and the data
(points). The dashed line shows the background
contribution.} 
\end{figure}

The results of our fits are presented in Table \ref{dpfit}. We find
that three fits (denoted as A-C) describe the data with similar
quality. The only difference among them is in description of the
$K^-\pi^+$ $S$-wave contribution, which is represented by the
$\overline{K}^*_0(1430)^0$ and NR in fit A, by
$\overline{K}^*_0(1430)^0$ and $\kappa(800)$ in fit B, and by the
LASS amplitude in fit C. Figure \ref{fig:lass} shows the Dalitz plot
for the $\kkp$ candidates and three projections of the data with the
result of fit B superimposed.

We generate seven sets of GEANT-based signal MC samples with the
model from fit A. Each set contains about the same size as in the
data. We find that the fits can recover the input magnitudes and
phases within their errors.

Fit B gives the best agreement with the data; thus we choose it to
search for $CP$-violation ($CPV$). The resonances in $D^+$ ($D^-$)
decays are allowed to have different magnitudes, $a_r + b_r$
($a_r-b_r$), and phases, $\delta_r + \phi_r$ ($\delta_r - \phi_r$),
in the decay amplitude ${\cal M}$ ($\overline{\cal M}$). We perform
a simultaneous fit to $D^+$ and $D^-$ samples. In the fit, the
signal term in Eq.~(\ref{lk}) is replaced by
\begin{equation}
{\cal L}_{\rm sig}=\frac{f\varepsilon^+(\msp,\msm)|{\cal
M}|^2}{\displaystyle \int \varepsilon^+(\msp,\msm)|{\cal
M}|^2\,d{\msp} d{\msm}}
\end{equation}
for the $D^+$ sample and by
\begin{equation}
\overline{\cal L}_{\rm
sig}=\frac{f\varepsilon^-(\msp,\msm)|\overline{\cal
M}|^2}{\displaystyle \int \varepsilon^-(\msp,\msm)|\overline{\cal
M}|^2\,d{\msp} d{\msm}}
\end{equation}
for the $D^-$ sample, where $\varepsilon^\pm$ are efficiency
functions obtained from the $D^\pm$ signal MC separately. We cannot
determine the relative magnitude and phase between $D^+$ and $D^-$
directly, and assume $b=0$ and $\phi=0$ for the $\overline{K}^{*0}$
resonance. The free parameters in the fit are $b_r/a_r$, $a_r$,
$\delta_r$, $\phi_r$ and $f$.

Following Ref.~\cite{ks2p}, we also compute the $CP$-conserving fit
fraction as
\begin{equation}
{\rm FF}(CPC)_r=\frac{\int|2a_r{\cal
A}_r|^2\,dm^2_+\,dm^2_-}{\int(|{\cal M}|^2+|\overline{\cal
M}|^2)\,dm^2_+\,dm^2_-}, \label{cpcff}
\end{equation}
the $CPV$ fit fraction as
\begin{equation}
{\rm FF}(CPV)_r=\frac{\int|2b_r{\cal
A}_r|^2\,dm^2_+\,dm^2_-}{\int(|{\cal M}|^2+|\overline{\cal
M}|^2)\,dm^2_+\,dm^2_-}, \label{cpvff}
\end{equation} and the $CPV$ interference fraction (IF) as
\begin{equation}
{\rm IF}_r=\frac{\left|\displaystyle\int \sum_{k\ne r} [2 a_k
e^{i\delta_k}
 \cos(\phi_k-\phi_r){\cal A}_k]\,b_r{\cal A}_r^*\,dm^2_+\,dm^2_-\right|}{\int(|{\cal
M}|^2+|\overline{\cal M}|^2)\,dm^2_+\,dm^2_-} \label{cpvif}.
\end{equation}
The $CP$-conserving fit fraction is the same for the $D^+$ and $D^-$
by construction. The $CPV$ fit fraction defined by Eq.~(\ref{cpvff})
is sensitive to $CP$ violation in the resonant decay. The $CPV$
interference fractions of Eq.~(\ref{cpvif}) sum over the
contribution proportional to $a_k e^{+i\delta_k}b_r$ so they are
sensitive to $CP$ violation in interference between resonances. The
phases are important and allow the possibility of cancelation in
this sum.

\begin{figure}[!hbtp]
\center
\includegraphics[width=0.9\textwidth]{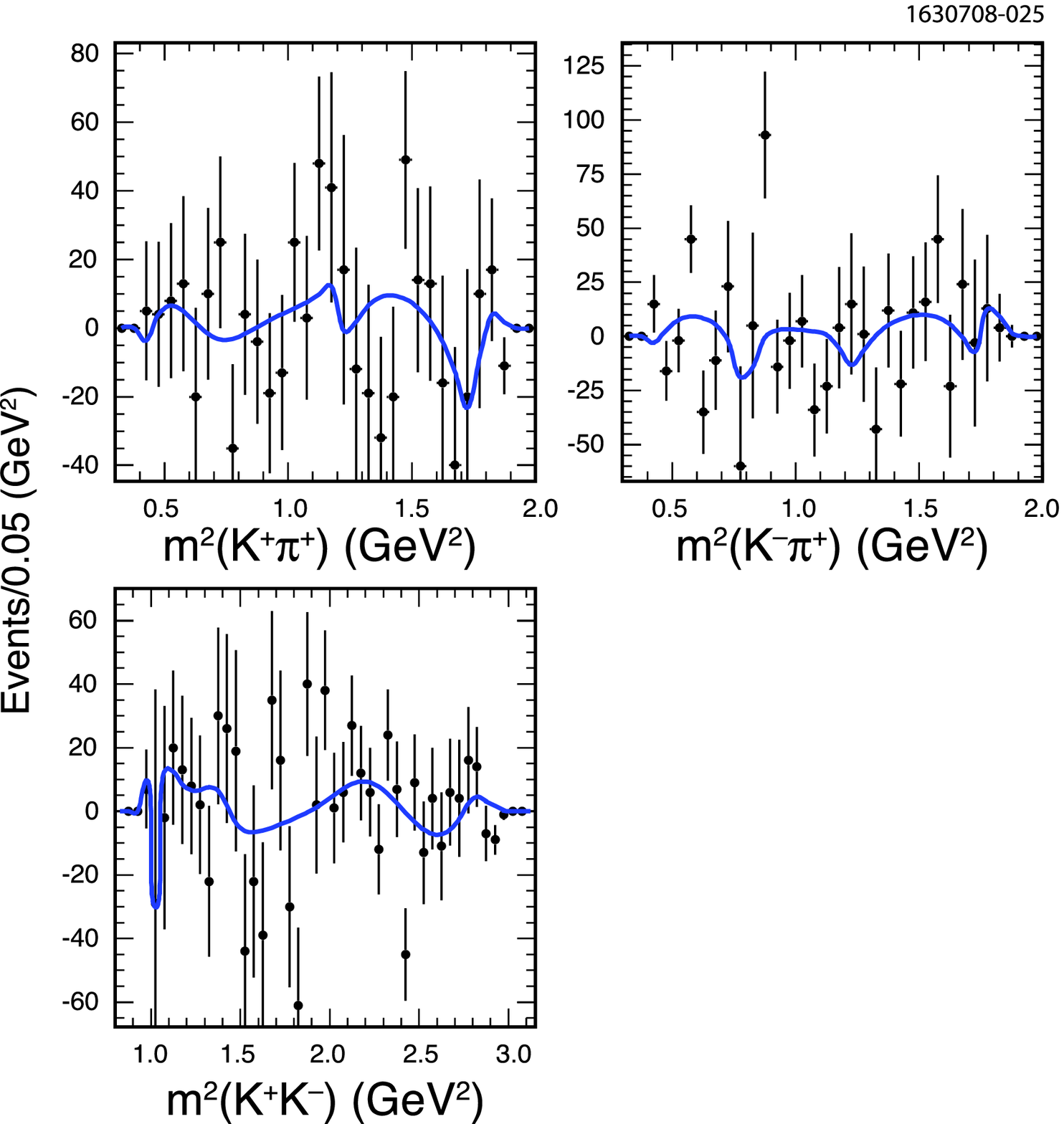}
\caption{\label{fig:dif} The difference of the Dalitz-plot
projections of data (points) and fit (line) between $\Dp$ and $\Dm$
decays.}
\end{figure}

In Table \ref{acp:lass}, we report the magnitude asymmetries
$b_r/a_r$, phase differences $\phi_r$ and fit fraction asymmetries.
The fit fraction asymmetry is computed as the difference between the
$D^+$ and $D^-$ fit fractions divided by the sum. The largest fit
fraction asymmetry, for the $\overline{K}_2^*(1430)^0$, is
1.7$\sigma$, and occurs because the fit fraction for the
$\overline{K}_2^*(1430)^0$ is small. The $CP$-conserving fit
fractions and the 95\% confidence level (C.L.) upper limits for
$CPV$ fit fraction, $CPV$ interference fraction, and the ratio of
$CPV$ interference to $CP$-conserving fit fraction are given in
Table \ref{t1}. We notice that the $CP$-conserving fit fractions are
consistent with those of fit B
in Table \ref{dpfit}. 
Figure \ref{fig:dif} shows the difference of the Dalitz-plot
projections of data and fit between $\Dp$ and $\Dm$ decays.

\begin{table}[!hbtp] \caption{\label{acp:lass}
The magnitude asymmetries $b_r/a_r$, phase differences $\phi_r$ and
 asymmetries on the $D^+$ and $D^-$ fit fractions from fit B. The errors are
statistical, experimental systematic, and decay-model systematic,
respectively.}
\begin{tabular}{cccc}\hline\hline
 $r$&$b/a$ (\%)&
$\phi$ ($^\circ$)&FF asymmetry(\%)\\\hline
$\overline{K}^{*0}$& 0(fixed)&0(fixed)&$-0.4\pm2.0_{-0.5-0.3}^{+0.2+0.6}$\\
$\overline{K}_0^*(1430)^0$&$4\pm3_{-0-1}^{+1+2}$&$-1\pm6_{-3-1}^{+0+6}$&
$8\pm6_{-1-1}^{+1+4}$\\
$\phi$&$-0.7\pm1.3_{-0.1-0.2}^{+0.2+0.3}$&$3\pm3_{-1-1}^{+0+3}$&
$-1.8\pm1.6_{-0.4-0.1}^{+0.0+0.2}$\\
$a_0(1450)^0$&$-10\pm7\pm2_{-3}^{+6}$&$4\pm3_{-2-1}^{+1+2}$&
$-19\pm12_{-3-11}^{+5+6}$\\
$\phi(1680)$&$-4\pm11_{-4-4}^{+5+6}$&$3\pm6\pm2_{-2}^{+3}$&
$-9\pm22_{-7-12}^{+10+9}$\\
$\overline{K}_2^*(1430)^0$&$23^{+12+1+3}_{-11-7-7}$&$5^{+5+1+3}_{-4-3-1}$&
$43\pm19_{-13-12}^{+1+5}$\\
$\kappa (800)$
&$-6\pm6_{-1-5}^{+3+1}$&$3\pm6_{-2-4}^{+4+1}$&$-12\pm11_{-6-2}^{+0+14}$\\\hline\hline
\end{tabular}
\end{table}

\begin{table}[!hbtp]
\center \caption{\label{t1} The $CP$-conserving fit fractions from
Eq.~(\ref{cpcff}) and the 95\% confidence level (C.L.) upper limits
for $CPV$ fit fraction from Eq.~(\ref{cpvff}), $CPV$ interference
fraction from Eq.~(\ref{cpvif}), and the ratio of $CPV$ interference
to
$CP$-conserving fit fraction. 
The 95\% C.L. upper limits include statistical and systematic
effects.}
\begin{tabular}{ccccc}\hline\hline
&&FF($CPV$) & IF  &Ratio\\
 &&  ($\times10^{-3}$)& ($\times10^{-3}$)&(\%)\\
 Component&FF($CPC$)(\%)&\multicolumn{3}{c}{(95\% C.L. upper limits)}
\\\hline
$\overline{K}^{*0}$&$25.7\pm0.5$&0(fixed)&0(fixed)&0(fixed)\\
$\overline{K}_0^*(1430)^0$&$18.8\pm1.2$&$<4.3$&$<12.6$&$<8.5$\\
$\phi$&$27.8\pm0.4$&$<0.6$&$<0.5$&$<0.17$\\
$a_0(1450)^0$&$4.7\pm0.6$&$<10.8$&$<31.6$&$<45$\\
$\phi(1680)$&$0.50\pm0.11$&$<0.9$&$<4.6$&$<89$\\
$\overline{K}_2^*(1430)^0$&$1.8\pm0.4$&$<6.9$&$<3.9$&$<22$\\
$\kappa (800)$&$7.0\pm0.8$&$<4.2$&$<17.2$&$<25$\\\hline\hline
\end{tabular}
\end{table}

We calculate an integrated $CP$ asymmetry across the Dalitz plot,
defined as 
\begin{equation}
 {\cal A}_{CP} = \int \frac{|{\cal M}|^2-|\overline{\cal
M}|^2}{|{\cal M}|^2+|\overline{\cal M}|^2} \,dm^2_+\,dm^2_- \Big /
\int \,dm^2_+\,dm^2_-. \label{ACP}
\end{equation}
We obtain ${\cal A}_{CP}=(-0.4\pm2.0_{-0.5-0.3}^{+0.2+0.6})\%$,
where the errors are statistical, experimental systematic, and
decay-model systematic, respectively.

Using the same counting technique as in Ref.~\cite{babar}, we
examine $CP$ asymmetries ($A_{CP}$) in the $\phi$ and
$\overline{K}^{*0}$ regions by requiring the $K^+K^-$ and $K^-\pi^+$
invariant mass to be within $15$ and $10$ MeV/$c^2$ of the nominal
$\phi$ and $\overline{K}^{*0}$ masses \cite{pdg}. We find $A_{CP}$
($-0.9\pm1.4\pm0.7$)\% and ($0.3\pm1.8\pm0.6$)\% for the $\phi$ and
$\overline{K}^{*0}$ region, respectively.

Systematic uncertainties from experimental sources and from the
decay model are considered separately. Our general procedure is to
change some aspect of our fit and interpret the change in the values
of the magnitudes, phases, fit fractions, $b_r/a_r$, $\phi_r$, and
fit fraction asymmetries as an estimation of the systematic
uncertainty.

Contributions to the experimental systematic uncertainties arise
from our model of the background, the efficiency and the event
selection. Our nominal fit fixes the coefficients of the background
determined from a sideband region. To estimate the systematic
uncertainty on this background shape, a fit is done with the
coefficients allowed to float and constrained by the covariance
matrix obtained from the background fit. Similarly, to estimate the
systematic uncertainty on the efficiency parameters, we perform a
fit with the coefficients of efficiency allowed to float constrained
by their covariance matrix. To estimate the systematic uncertainty
on MC simulation for the particle identification, a fit is done with
new efficiency parameters obtained from the weighted MC sample by
the efficiency ratios of data to MC depending on each particle's
momentum. To estimate the event selection uncertainty, we change the
$\de$ and $\mbc$ selection criteria in the analysis. These
variations to the standard fit are the largest contribution to our
experimental systematic errors. In the $CP$ asymmetry search, we
take the background fractions and shapes to be the same for the
$D^+$ and $D^-$ samples. To estimate the uncertainty on the
supposition, we perform a fit with the background determined
separately.

The systematic error due to our choice of $\kkp$ decay model is
evaluated as follows.  We change the standard values of the radial
parameter in the Blatt-Weisskopf form factors \cite{bwform} for the
intermediate resonance decay vertex (1.5 GeV$^{-1}$) and the $D^+$
vertex (5 GeV$^{-1})$ both to 1 GeV$^{-1}$.  Fits with constant
width in the Breit-Wigner functions are considered. To compute the
uncertainty arising from our choice of resonances included in the
fit, we compare the result of our standard fit to a series of fits
where each of the resonances, $\overline{K}^*(1410)^{0}$,
$f_0(980)$, $f_0(1370)$, $f_0(1500)$, $f_2(1270)$,
$f_2^\prime(1525)$, $a_0(980)^0$ and $a_2(1320)^0$, is included one
at a time. These variations to the standard fit result in the
largest contribution to systematic errors associated with our decay
model. The masses and widths of the intermediate resonances are
allowed to vary within their known uncertainties \cite{pdg}. For fit
C, we vary the parameters in the LASS amplitude within their
uncertainties.

We take the maximum variation of the magnitudes, phases, and fit
fractions, $b_r/a_r$, $\phi_r$, and fit fraction asymmetries from
the nominal fit compared to the results in this series of fits as a
measure of the experimental systematic and decay-model systematic
uncertainty. Table \ref{sys-acp} shows the systematic checks on the
integrated $CP$ asymmetry defined in Eq.~(\ref{ACP}). Apart from the
sources discussed above, we also consider different models from fit
A or C; the variations are small.

\begin{table}[!hbtp]
\center\caption{\label{sys-acp} Sources contributing to systematic
uncertainties on the integrated $CP$ asymmetry defined in
Eq.~(\ref{ACP}).}
\begin{tabular}{lccc}\hline\hline
Source& Variation (\%)\\\hline
Background shape&$-0.01$\\
Efficiency parameters&$+0.02$\\
Particle identification&$+0.06$\\
Event selection criteria& $+0.18$\\
Background (in)dependent fit& $-0.52$\\
\hline
Form factors&$+0.21$\\
Width parameterization&$-0.15$\\
Choice of resonances&$^{+0.61}_{-0.33}$\\
Resonant masses and widths&$^{+0.09}_{-0.08}$\\
Fit A&$+0.07$\\
Fit C&$-0.15$
\\\hline\hline
\end{tabular}
\end{table}

We estimate the systematic uncertainty on the $CP$ asymmetry defined
in Eq.~(\ref{acp_yield}). The contributions from various identified
sources are listed in Table \ref{syscp}. The uncertainty due to
selection criteria is estimated by doubling the $\de$ signal window.
We evaluate an uncertainty for the background shape by floating its
parameters in the fit instead of fixing them from the values
obtained form the $\de$ sideband. We use the $CP$-conserved channels
$D^+\to K^-\pi^+\pi^+$ and $D^0\to K^-\pi^+ \pi^0$ as control modes
to assign the systematic uncertainty on MC simulation due to
possible efficiency difference on positive and negative charged
kaons and pions.

\begin{table}[!hbtp]
\center\caption{\label{syscp} Systematic uncertainties on the $CP$
asymmetry defined in Eq.~(\ref{acp_yield}).}
\begin{tabular}{cccc}\\\hline\hline
Source&Variation (\%)\\
\hline
Selection criteria &$\pm0.25$\\

Background shape& $\pm0.02$ \\

MC simulation& $\pm0.15$\\\hline
 Total& $\pm0.29$\\\hline\hline
\end{tabular}
\end{table}

In conclusion, we have analyzed the resonant substructure in $D^+
\to K^+K^-\pi^+$ decay and searched for $CP$ violation in the decay
and its intermediate resonances. We measure the overall $CP$
asymmetry in $D^\pm\to K^+K^-\pi^\pm$ decays 
to be
$(-0.03\pm0.84\pm 0.29)\%$. The limit is more restrictive than the
one found previously by \babar \cite{babar}. We use five resonances
and $K^-\pi^+$ $S$-wave states to model the Dalitz plot with results
shown in Table \ref{dpfit}. The $K^-\pi^+$ $S$-wave can be equally
well described by a coherent sum of $\overline{K}_0^*(1430)^0$ and
nonresonant amplitude or $\overline{K}_0^*(1430)^0$ and
$\kappa(800)$, or the LASS amplitude. Choosing the second model 
we measure the $CP$ asymmetries for all submodes, shown in Table
\ref{acp:lass} and \ref{t1}. The measured $CP$ asymmetries are
consistent with the absence of $CP$ violation. We find ${\cal
A}_{CP}$ defined in Eq.~(\ref{ACP}) to be
$(-0.4\pm2.0_{-0.5-0.3}^{+0.2+0.6})\%$. The ${\cal A}_{CP}$ is
sensitive to an asymmetry in shape between the $D^+$ and $D^-$
samples, but does not depend on their yields.

We gratefully acknowledge the effort of the CESR staff in providing
us with excellent luminosity and running conditions.
D.~Cronin-Hennessy and A.~Ryd thank the A.P.~Sloan Foundation. This
work was supported by the National Science Foundation, the U.S.
Department of Energy, the Natural Sciences and Engineering Research
Council of Canada, and the U.K. Science and Technology Facilities
Council.

\end{document}